\documentclass[prb,showpacs]{revtex4}
\usepackage{graphicx}% Include figure files
\usepackage{bm}% bold math

\begin{document}

\title{Optical excitation of interacting
    electron-hole pairs in disordered one-dimensional semiconductors}
\author{Maxim Mostovoy, Frank Antonsen, and Jasper
Knoester} \affiliation{Institute for Theoretical Physics,
Materials Science Center\\ Rijksuniversiteit Groningen\\
Nijenborgh 4, 9747 AG Groningen\\
  The Netherlands}
\date{\today}


\begin{abstract}

We apply the optimal fluctuation method to the calculation
of the optical absorption in disordered one-dimensional
semiconductors below the fundamental optical gap. We find
that a photon energy exists at which the shape of the
optimal fluctuation undergoes a dramatic change, resulting
in a different energy dependence of the absorption rate
above and below this energy. In the limit when the
interaction of an electron and a hole with disorder is
stronger than their interaction with each other, we obtain
an analytical expression for the optical conductivity. We
show that to calculate the absorption rate, it is, in
general, necessary to consider a manifold of optimal
fluctuations, rather than just a single fluctuation. For
an arbitrary ratio of the Coulomb interaction and
disorder, the optimal fluctuation is found numerically.

\end{abstract}

\maketitle

\section{\label{intro}Introduction}

The interplay between disorder and inter-particle
interactions results in a number of remarkable phenomena,
e.g., a singularity in the electron density of states at
the Fermi
energy\cite{AltshulerAronov,AAL,EfrosShklovskii,Efros} and
an enhanced localization length of pairs of interacting
particles.\cite{tip} A familiar situation in which the
simultaneous presence of Coulomb interactions and disorder
plays an important role, is the process of optical
absorption in disordered semiconductors below the
fundamental optical gap, i.e., below the band to band
transition.\cite{MottDavis}  With the advent of modern
optical materials, this problem is not only of interest in
its three-dimensional version, for which it has received
most attention, but also in two dimensions (quantum
wells)\cite{Haug,Ogawa} and one dimension (quantum
wires\cite{Haug,Ogawa} and semiconducting
polymers\cite{Hadzii}).  Interestingly, in the absorption
spectrum the relative strength of the Coulomb interaction
between electron and hole and their interactions with the
disorder, does not only depend on the disorder strength,
but also on the photon energy: the lower the photon
energy, the larger the amplitude of a disorder fluctuation
should be in order to create the corresponding absorbing
state below the fundamental optical gap.

The effect of a relatively weak disorder on the Wannier
exciton was considered in Refs.
\onlinecite{BaranovskiiEfros} and \onlinecite{EfrosRaikh}.
Here, "relatively weak" refers to the situation where the
exciton absorption peak is still visible as a separate
peak below the band-to-band transition and is not smeared
entirely by the disorder. Moreover, the restriction to
weak disorder only applies for photon energies close to
the exciton energy in the absence of disorder, $E_{ex}$.
In this case, the exciton localization length is much
larger than the average electron-hole separation, and the
exciton center-of-mass motion decouples from the relative
motion of the electron and hole.  The Wannier exciton then
essentially behaves as a Frenkel exciton in an effective
disorder potential and the calculation of the optical
absorption spectrum becomes a single-particle problem.
This effective approach has been used to study numerically
the absorption and luminescence line shapes due to
excitons in semi-conductor quantum wells with interface
roughness.\cite{Zimm}

In Refs.~\onlinecite{BaranovskiiEfros} and
\onlinecite{EfrosRaikh}, the low-energy tail of the
exciton absorption peak was calculated analytically within
the effective one-particle approach, by using the optimal
fluctuation method.\cite{hl,zl} This method applies when
the dominant contribution to the quantity of interest
comes from disorder realizations close to a single large
disorder fluctuation. This method can be formulated as a
saddle-point calculation of the functional integration
over disorder realizations.\cite{lif} It was used
previously to calculate the optical absorption in
disordered Peierls conductors\cite{MK} and it is similar
to the calculation of the Urbach tails resulting from the
interaction of the electron-hole pair with the lattice,
treated in the quasistatic approximation.\cite{Ioselevich}

In this paper we apply the optimal fluctuation method to
the general case of arbitrary ratio of disorder strength
and Coulomb interaction. In the strong disorder limit,
when the exciton peak is destroyed, and for weak disorder,
assuming sufficiently low photon energies, we obtain an
analytical expression for the optical conductivity. We
show that in these situations, the electron and hole in
the optimal fluctuation are localized in two separate
potential wells. We also show that in order to calculate
the optical absorption rate, one has, in general, to
consider a manifold of optimal fluctuations with different
values of the electron-hole separation, rather than just
one fluctuation.

In the general case of arbitrary ratio of the Coulomb
interaction and disorder strength the equation for the
optimal fluctuation is solved numerically. We do this for
an electron and hole described within the one-dimensional
tight-binding model. The equation for the optimal
fluctuation in this model is similar to the equation for
polaronic excitons and bipolarons obtained in the
adiabatic approximation.\cite{EminHolstein,Emin,Aubry} We
show that, at some energy $E_c$, lying in the region where
the Coulomb interaction and disorder are of the same
order, the shape of the optimal fluctuation undergoes a
sudden change: while above $E_c$ it has the form of a
single potential well, below $E_c$ it consists of a
``dip'' that localizes the hole and a ``bump'' that
localizes the electron. This cross-over only takes place
in one-dimensional systems.

This paper is organized as follows. In Sec. \ref{conmod}
we introduce the continuum model that describes
interacting electron-hole pairs in disordered
semiconductors. To make the reader more familiar with the
optimal fluctuation method, in Sec.~\ref{1particle} we
briefly discuss its application to the calculation of the
one-particle density of states. Then, in
Sec.~\ref{optfluc}, we will discuss the shape of the
optimal fluctuation for the electron-hole states. We argue
that this shape crucially depends on the dimensionality of
the system.  We obtain and solve perturbatively the
nonlinear nonlocal equation for the optimal fluctuation in
the one-dimensional case. In Sec.~\ref{optabs} we obtain
an expression for the tails of the optical absorption
spectrum, which has a wider range of validity than the
optimal fluctuation method. In Sec.~\ref{numres} we
present and discuss our results for the optimal
fluctuation. Finally, we summarize and conclude in
Sec.~\ref{conclude}. Some technical details have been
moved to appendices in order not to disturb the natural
flow of the text.

\section{\label{conmod}Continuum model and optical absorption}

We consider direct gap semiconductors with Wannier
excitons, in which case the low-energy electron-hole
states can be described in the continuum approximation. In
this approximation, the wave function $\Psi_\alpha({\bf
x}_1,{\bf x}_2)$ of the electron-hole pair with the energy
$E_\alpha$, counted from the gap value $\Delta$, satisfies
the Schr\"odinger equation:
\begin{equation}
\left[
  -\frac{\hbar^2}{2m_h} \triangle_1 -
   \frac{\hbar^2}{2m_e} \triangle_2
  + U({\bf x}_1) - \beta U({\bf x}_2)
  + V({\bf x}_1 - {\bf x}_2)
\right] \Psi_\alpha = E_\alpha \Psi_\alpha. \label{Sch}
\end{equation}
Here, ${\bf x}_{1}$ and  ${\bf x}_{2}$ are the coordinates
of, respectively, hole and electron, $m_h$($m_e$) is the
effective hole(electron) mass, $V({\bf x}) = -
\frac{e^2}{\epsilon x}$ is the Coulomb interaction between
the electron and hole, $U({\bf x})$ is the random
potential due to impurities acting on the hole, while the
disorder potential acting on the electron is $- \beta
U({\bf x})$. The dimensionless coefficient $\beta$
accounts for a different dependence of the energies of the
bottom of the conduction band and the top of the valence
band on the concentration of impurities.\cite{EfrosRaikh}
We consider here the case of uncorrelated white noise
disorder of strength $A$:
\begin{equation}
\langle U({\bf x}) U({\bf x}^\prime) \rangle = A
\delta({\bf x} - {\bf x}^\prime). \label{whitenoise}
\end{equation}

The optical conductivity per unit volume for the electric
field, polarized, {\em e.g.}, in the $x$-direction, is
given by
\begin{equation}
\sigma(\omega) = \frac{C}{\omega} F(\hbar \omega -
\Delta), \label{sigma}
\end{equation}
where the coefficient $C$,
\begin{equation}
C = \frac{2 \pi e^2 \hbar^2}{m^2} \, \left|\int_{V_0} d^3r
u_{0}^{\ast}({\bf r}) \frac{\partial}{\partial
x}v_{0}({\bf r})\right|^2, \label{C}
\end{equation}
is expressed through the periodic Bloch waves, $u_0({\bf
r})$ and $v_0({\bf r})$, describing, respectively, the
electron and hole states with zero wave vector. In
Eq.(\ref{C}) $m$ is the electron mass in vacuum and the
integration goes over one elementary unit cell $V_0$, in
which the functions $u_0({\bf r})$ and $v_0({\bf r})$ are
normalized to unity.

The function $F$ in Eq.(\ref{sigma}) is the part of the
optical conductivity that has to be calculated within the
continuum model:
\begin{equation}\label{F}
F(\hbar \omega - \Delta) = \frac{1}{V} \left\langle
\sum_\alpha \left| D_{\alpha0} \right|^2
\delta\left(\Delta + E_\alpha - \hbar \omega
\right)\right\rangle.
\end{equation}
In the last equation
\begin{equation}\label{D}
D_{\alpha 0} = \int d^dx \Psi_{\alpha}^{\ast}({\bf x},{\bf
x}),
\end{equation}
is the ``continuum part'' of the matrix element of the
transition from the ground state to the excited state
$\alpha$ (see Eq.(\ref{Sch})). In Eq.(\ref{F}) the
brackets $\langle \ldots \rangle$ denote the disorder
average, $V = L^d$ is the total volume, $L$ is the linear
size, and $d$ is the dimensionality of the system.

\section{\label{1particle}The One-Particle Case}

In this section we briefly recall how the optimal
fluctuation method can be used to calculate the low-energy
tail of the density states (per unit
volume),\cite{hl,zl,lif}
\begin{equation}
\rho(\varepsilon) = \frac{1}{V}\left\langle \sum_\alpha
\delta\left(\varepsilon_\alpha[U] -
\varepsilon\right)\right\rangle, \label{rho1}
\end{equation}
of a single particle moving in a random potential:
 \begin{equation}
      H \psi_\alpha =
      \left(-\frac{\hbar^2}{2m}\triangle + U({\bf x}) \right)
      \psi_\alpha =
      \varepsilon_\alpha\psi_\alpha,
      \;\;\alpha=0,1,2,\ldots\label{eq:schr1}
\end{equation}
We are now interested in the density of states with a
large negative energy. Such states can only be induced by
large negative fluctuations of the disorder potential
$U(x)$ (in the absence of disorder the energy of all
eigenstates is positive). The density of state is then,
essentially, the probability to find such a fluctuation.
When this probability is small, it suffices to keep in
Eq.(\ref{rho1}) only the contribution of the ground state
($\alpha = 0$):
\begin{equation}
\rho(\varepsilon) \approx \frac{1}{V}\left\langle
\delta\left(\varepsilon_0[U] -
\varepsilon\right)\right\rangle, \label{rho2}
\end{equation}
because the probability to find a disorder fluctuation
that induces an excited state with the same energy is even
smaller.

For the white noise potential Eq.(\ref{whitenoise}) the
disorder average can be performed by functional
integration:
\begin{eqnarray}
\rho(\varepsilon) &\!\!\!=\!\!\!& \frac{1}{V}\int\!\!{\cal
D}U e^{-\frac{S}{A}} \delta\left(\varepsilon_0[U] -
\varepsilon\right) \nonumber\\ &\!\!\!=\!\!\!&
\frac{1}{V}\int\!\!{\cal D}U \frac{d\lambda}{2\pi i A}
e^{-\frac{1}{A}\left(S +
\lambda(\varepsilon_0[U]-\varepsilon)\right)},
\label{funint}
\end{eqnarray}
where the ``action'' $S$ is given by
\begin{equation}
S = \frac{1}{2} \int\!\!d^dx U^2({\bf x}).
\label{defaction}
\end{equation}

The optimal fluctuation method is the saddle-point
calculation of the functional integral in
Eq.(\ref{funint}), in which one assumes that the dominant
contribution to this integral comes from the vicinity of
one ``optimal'' disorder fluctuation $U({\bf x})$, which
has the highest weight among the disorder realizations
that induce a state at the energy $\varepsilon$. At the
``saddle-point'' the variation of
\begin{equation}
S_\lambda = S + \lambda(\varepsilon_0[U]-\varepsilon)
\label{Slam1}
\end{equation}
with respect to $U({\bf x})$ vanishes:
\begin{equation}
        0=\frac{\delta S_\lambda}{\delta U({\bf x})} = U({\bf x}) +\lambda
        \frac{\delta \varepsilon_0[U]}{\delta U({\bf x})}.
\end{equation}
The Feynman-Hellmann theorem allows us the carry out the
variation of the energy with respect to the disorder as
\begin{equation}
        \frac{\delta \varepsilon_0[U]}{\delta U({\bf x})} =
        \langle 0 | \frac{\delta H}{\delta U({\bf x})}
        | 0 \rangle = |\psi_0({\bf x})|^2,
          \label{eq:FH}
\end{equation}
whence
\begin{equation}
          U({\bf x}) = -\lambda |\psi_0({\bf x})|^2 . \label{eq:optU}
\end{equation}
We must then insert this into the original Schr\"{o}dinger
equation (\ref{eq:schr1}) to find $\psi_0$. The resulting
effective equation for $\psi_0$ (which we can take to be
real) is then the nonlinear Schr\"{o}dinger equation
\begin{equation}
\frac{\hbar^2}{2m}\triangle \psi_0 + \lambda \psi_0^3 +
\varepsilon \psi_0 = 0 \label{eq:nlse}.
\end{equation}

It is convenient to introduce the dimensionless coordinate
${\bf z} = \kappa {\bf x}$, where $\kappa$ is defined by
$\varepsilon = - \frac{\hbar^2 \kappa^2}{2m}$ and the
dimensionless wave function $\phi({\bf z})$:
\begin{equation}
\psi_0({\bf x}) = \frac{\hbar \kappa}{\sqrt{\lambda
m}}\phi({\bf z}), \label{psiphi}
\end{equation}
which satisfies
\begin{equation}
\triangle_{\bf z} \phi + 2 \phi^3 - \phi = 0.
\label{nonlin}
\end{equation}
In dimensionless units the spatial extent of the wave
function and the optimal fluctuation is of the order of
$1$. For $d = 1$, the solution of Eq.(\ref{nonlin}) can be
easily found analytically
\begin{equation}
\phi(z) = \frac{1}{\cosh z}, \label{phi}
\end{equation}
while for $d > 1$, this equation has to be solved
numerically.

The solution of Eq.(\ref{nonlin}) is an extremum of the
functional
\begin{equation}
{\cal A}[\phi] = \int\!\!d^dz \left[\left(\nabla_{\bf z}
\phi \right)^2 + \phi^2 -\phi^4 \right] \equiv a - b + c,
\label{A}
\end{equation}
where we have used the notations
\begin{equation}
\left\{
\begin{array}{rcl}
a &=& \int\!\!d^dz \phi^2,\\ \\b &=& \int\!\!d^dz
\phi^4,\\
\\ c &=& \int\!\!d^dz \left(\nabla_{\bf z} \phi \right)^2.
\end{array}
\right.
\end{equation}
The parameters of the optimal fluctuation can be expressed
through $a$ and $b$. Thus, the coefficient $\lambda$ in
Eq.(\ref{eq:optU}), found using the normalization
condition $\int\!\!d^d x \psi_0^2({\bf x}) = 1$ and
Eq.(\ref{psiphi}), is expressed through $a$ by
\[
\lambda = a \frac{\hbar^2 \kappa^{2-d}}{m}
\]
and the action $S$, defined by Eq.(\ref{defaction}), for
the optimal fluctuation is given by
\begin{equation}
S = 2^{1 - d/2} b \,\frac{\hbar^d |\varepsilon|^{2 -
d/2}}{m^{d/2}}. \label{action1}
\end{equation}

One can obtain relations between the coefficients $a$,
$b$, and $c$, using scaling arguments. In particular, if
$\phi({\bf z})$ is a solution of Eq.(\ref{nonlin}), then
for the rescaled function $\phi_\Lambda({\bf z}) = \Lambda
\phi({\bf z})$, we have
\begin{equation}
\left.\frac{d}{d\Lambda} {\cal A}[\phi_\Lambda]
\right|_{\Lambda = 1} = 0. \label{dLam}
\end{equation}
Since ${\cal A}[\phi_\Lambda] = \Lambda^2(a+c) - \Lambda^4
b$, we obtain
\begin{equation}
a+c = 2b. \label{rel1}
\end{equation}
Furthermore, the rescaling of coordinates,
$\phi_\Lambda({\bf z}) = \phi(\Lambda^{-1}{\bf z})$ upon
substitution into Eq.(\ref{dLam}) gives
\begin{equation}
(d-2)c + d(a-b) = 0. \label{rel2}
\end{equation}
Combining Eqs.(\ref{rel1}) and (\ref{rel2}), we obtain
\[
\left\{
\begin{array}{rcl}
b &=& \frac{2}{(4-d)} a,\\ \\c &=& \frac{d}{(4-d)}a.
\end{array}
\right.
\]

We note that the same scaling arguments were used to prove
that a scalar theory cannot have an inhomogeneous
classical minimum in dimensions greater than 1 (the
so-called Derrick's theorem\cite{Derrick}). Nonetheless,
the inhomogeneous solution $\phi({\bf z})$ of
Eq.(\ref{nonlin}) exists both for $d=2$ and $d=3$, as it
gives an extremum of ${\cal A}$ (that is not bounded from
below) rather than its minumum.

\begin{figure}[htbp]
  \begin{center}
    \includegraphics[width=7cm]{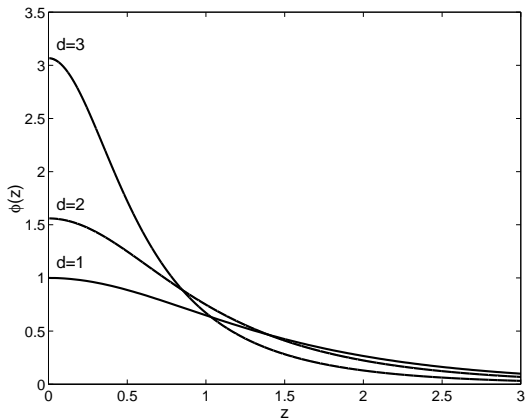}
    \caption{The shape of the wave function $\phi(z)$ for $d =1,2,3$.}
    \label{phid}
  \end{center}
\end{figure}

The dependence of the function $\phi$ on the radius $z =
|{\bf z}|$ is shown in Fig.~\ref{phid}. For $d =2,3$ we
solved Eq.(\ref{nonlin}) numerically (see {\em e.g.},
Refs.~\onlinecite{BrezinParisi} and \onlinecite{Parisi})
while for $d = 1$ we used the analytical solution
(\ref{phi}) for $z>0$. The value of the coefficient $a$ is
\[
a = \left\{
\begin{array}{cll}
2, &\mbox{for}&d=1,\\ \\ 5.85\ldots, &\mbox{for}&d=2,\\
\\6.30\ldots, &\mbox{for}&d=3.
\end{array}
\right.
\]

The result for the single-particle density of states at
large negative energies, obtained by the optimal
fluctuation method, has the form
\begin{equation}
     \rho(\varepsilon) = K(\varepsilon)
     e^{-\frac{S}{A}} \label{eq:rho1},
\end{equation}
where the prefactor $K(\varepsilon)$ results from the
Gaussian integrations over the small deviations of $\delta
U(x)$ from the optimal fluctuation. The calculation of the
prefactor also involves the integration over the locations
of the optimal fluctuation, which cancels the volume $V$
in the denominator of Eq.(\ref{rho1}). For $d = 1$ the
prefactor $K(\varepsilon)$ is given by\cite{hl}
\begin{equation}
K(\varepsilon) = \frac{4 |\varepsilon|}{\pi A }.
\label{K1}
\end{equation}
An alternative derivation of this result, in which we use
the correspondence between the optimal fluctuation and the
instanton, describing the tunneling in the double-well
problem, is given in Appendix~A.

\section{\label{optfluc}Optimal fluctuation for electron-hole states}

We now return to the problem of an interacting
electron-hole pair in the presence of disorder. Similar to
the previous section, we can obtain an equation for the
wave function of the typical electron-hole state with
large negative energy. However, before turning to a formal
consideration (see Eqs. (\ref{Psi0}) and further), we
first give a qualitative discussion of the properties of
the optimal disorder fluctuation for the excitonic states.

From Eq.(\ref{eq:optU}) we see, that the spatial extent of
the disorder fluctuation that induces the typical
single-particle state with the negative energy
$\varepsilon$, equals the spatial extent of this state,
given by $\kappa^{-1} \propto |\varepsilon|^{-1/2}$.
Similarly, for the electron-hole state with negative
energy $E$, the spatial extent of the optimal fluctuation,
$r \propto|E|^{-1/2}$. While the magnitude of the disorder
potential $\propto |E|$, the magnitude of the Coulomb
energy of the electron-hole pair $\propto r^{-1} \propto
|E|^{1/2}$. Thus, for large negative $E$, the Coulomb
energy becomes smaller than the energy of the interaction
of the electron and hole with disorder and the Coulomb
interaction can be treated perturbatively. This situation
is very similar to that of the Coulomb gas, which becomes
more ideal as its density increases.\cite{Landau}

The shape of the typical electron-hole state crucially
depends on the dimensionality of the system $d$. We first
neglect the electron-hole interaction completely and
consider a localized hole with the energy
$\varepsilon_h<0$ and a localized electron with the energy
$\varepsilon_e < 0$, such that $\varepsilon_h +
\varepsilon_e = E$. Assuming that the electron and hole
are localized far from each other, the action of the
corresponding disorder fluctuation is given by the sum of
the single-particle actions (see Eq.(\ref{action1})):
\begin{equation}
S = S_h(\varepsilon_h) + S_e(\varepsilon_e) = 2^{1 - d/2}
b \hbar^d \left(\frac{|\varepsilon_h|^{2 -
d/2}}{m_h^{d/2}} + \frac{|\varepsilon_e|^{2 -
d/2}}{\beta^2 m_e^{d/2}} \right). \label{action2}
\end{equation}
The factor $\beta^2$ in the electron action is due to the
fact that the strength  of the disorder potential, acting
on the electron, is $\beta^2 A$.

For $d = 3$ the minimum of the action is reached when only
one particle is localized, {\em i.e.} at
\[
\left\{
\begin{array}{rl}
\varepsilon_h = E, \; \varepsilon_e = 0,  & \mbox{for}
\;\;m_h
> m_e \beta^{4/3},\\ \\ \varepsilon_h = 0, \;\varepsilon_e = E,
& \mbox{for} \;\;m_h < m_e \beta^{4/3}.
\end{array}
\right.
\]

If we now include the Coulomb interaction between the
electron and hole, we obtain the following picture of the
typical electron-hole state with a large negative energy:
One particle is localized by a disorder fluctuation, while
the other particle forms a bound hydrogen-like state with
the localized particle. If {\em e.g.} the hole is
localized, then the total pair energy is
\begin{equation}
E \simeq \varepsilon^{(1)} + \varepsilon^{(2)} +
\varepsilon^{(3)} \label{totalE}
\end{equation}
where $\varepsilon^{(1)} = - \frac{\hbar^2
\kappa_{h}^{2}}{2 m_h}$ is the energy of the localized
hole, the second term is the binding energy of the
electron, $\varepsilon^{(2)} = - \frac{m_{e}
e^4}{2{\epsilon}^2 \hbar^2}$, and the third term is the
energy of the repulsion of the electron from the disorder
potential that localizes the hole. This picture is valid
when $|\varepsilon^{(2)}| \ll |\varepsilon^{(1)}|$, in
which case the size of the hydrogen-like state, $r =
\epsilon \frac{m}{m_e} a_B$ (here $a_B$ is the Bohr
radius), is much larger than spatial extent of the
disorder fluctuation localizing the hole $\kappa_h^{-1}$:
\[
\kappa_h r = \sqrt{\frac{m_h \varepsilon^{(1)}}{m_e
\varepsilon^{(2)}}} \gg 1.
\]
In this limit the disorder potential acting on the
electron can be approximated by $\delta-$function and the
corresponding energy, $\varepsilon^{(3)} \simeq 50
\left(\frac{m_e}{m_h}\right)^{3/2}
\frac{|\varepsilon^{(2)}|^{3/2}}{|E|}$, is parametrically
smaller than $|\varepsilon^{(2)}|$ and can be neglected.
Then the absorption rate is:
\[
F(E) \propto e^{-\frac{S_h}{A}} = \exp\left\{-
\frac{b\hbar^3}{2^{1/2}m_h^{3/2}A} \left| E + \frac{m_{e}
e^4}{2\epsilon^2\hbar^2}\right|^{1/2}\right\}.
\]

For $d = 2$, the situation is, essentially, the same as
for $d = 3$: the minimization of the action
(\ref{action2}) gives that in the optimal fluctuation only
one of the two particles is localized by disorder (the
hole, for $m_h
> \beta^2 m_e$, or the electron, otherwise). For $d = 1$
the situation is quite different. In that case the minimum
of the action (\ref{action2}) is reached at
\begin{equation}
\frac{\varepsilon_e}{\varepsilon_h} = \beta^4
\frac{m_e}{m_h}, \label{ratio}
\end{equation}
which means that in the one-dimensional electron-hole
state with large negative energy both the electron and the
hole are likely to be localized. In that case the optimal
fluctuation has two parts (cf. Figs.~\ref{fig:optflu}(b)
and \ref{fig:optflu}(c) below): a part where the disorder
potential is negative (to localize the hole) and a part
with positive disorder potential (to localize the
electron). The minimal value of the action is then given
by:
\begin{equation}
S_0 = \frac{4\hbar}{3} \sqrt{ \frac{2|E|^3}{M}},
\label{S0}
\end{equation}
where $M = m_h + \beta^4 m_e$.

There are two kinds of corrections to this action: (i) the
correction due to the electron-hole interaction and (ii)
the correction due to the interaction of the electron with
the disorder fluctuation localizing the hole and {\em vice
versa}. For large negative $E$ these corrections are
relatively small. The regular way to find them is the
perturbative solution of the equation for the typical wave
function of the electron-hole state of a given energy.
This equation can be obtained using the same
considerations that led to Eq.(\ref{eq:nlse}). The
relation between the optimal fluctuation $U({\bf x})$ and
the wave function of the electron-hole state $\Psi({\bf
x}_1,{\bf x}_2)$ it induces, has the form
\begin{equation}
U({\bf x}) = -\lambda \int d^dx' \left(\Psi^2({\bf x},{\bf
x}') - \beta \Psi^2({\bf x}',{\bf x})\right), \label{U2}
\end{equation}
where we used that $\Psi$ satisfies Eq.(\ref{Sch}).
Substituting this optimal fluctuation back into the
Schr\"{o}dinger equation (\ref{Sch}) leads to a nonlinear,
nonlocal equation for $\Psi$:
\begin{eqnarray}
\bigg(\!\!
  &\!\!\!-\!\!\!&\frac{\hbar^2}{2m_h} \triangle_1
  -\frac{\hbar^2}{2m_e} \triangle_2
  +V({\bf x}_2-{\bf x}_1)-E\bigg)\Psi({\bf x}_1,{\bf x}_2)
  = \lambda \Psi({\bf x}_1,{\bf x}_2) \nonumber \\
  &\!\!\!\times\!\!\!&\!\!\int\!\!d^dx'
  \left(\Psi^2({\bf x}_1,{\bf x}')
-\beta \Psi^2({\bf x}',{\bf x}_1) -\beta \Psi^2({\bf
x}_2,{\bf x}') + \beta^2\Psi^2({\bf x}',{\bf x}_2)\right)
     \label{eq:nonloc} .
\end{eqnarray}
This equation is formally equivalent to the equation for
the excitonic polaron and bipolaron wave functions,
obtained in the adiabatic treatment of the
lattice.\cite{EminHolstein,Emin,Aubry} It is clearly
impossible to solve this equation analytically. In the
remainder of this section we obtain the action of the
optimal fluctuation for $d =1$ using a perturbative
solution of this equation, while in Sec.~\ref{numres} we
give the results of a numerical solution, also for $d =
1$.

The first term in the perturbative expansion of the wave
function,
\[
\Psi = \Psi_0 + \Psi_1 + \ldots,
\]
is the product of the wave functions of a noninteracting
electron-hole pair separated by some distance:
\begin{equation}
\Psi_0(x_1,x_2) = \psi_h(x_1-x_h)\psi_e(x_2 - x_e) =
\frac{\sqrt{\kappa_h \kappa_e}}{2} \phi(\kappa_h (x_1 -
x_h))\phi(\kappa_e(x_1 - x_e)), \label{Psi0}
\end{equation}
where $x_h$($x_e$) are the average hole(electron)
positions, the single-particle wave function $\phi$ is
given by Eq.(\ref{phi}) and the wave vectors $\kappa_h$
and $\kappa_e$ are defined by $\varepsilon_h = -
\frac{\hbar^2 \kappa_h^2}{2m_h}$ and $\varepsilon_e = -
\frac{\hbar^2 \kappa_e^2}{2m_e}$.

In the zeroth order of the expansion the disorder
potential in the optimal fluctuation is obtained by
substituting Eq.(\ref{Psi0}) into Eq.(\ref{U2}):
\begin{equation}
U_0(x) = U_h(x - x_h) + U_e(x - x_e) = - \lambda
\psi_h^2(x-x_h) + \beta \lambda \psi_e^2(x - x_e),
\label{U0}
\end{equation}
where the first term is the negative potential (a dip)
that localizes the hole and the second term is the
positive potential (a bump) that localizes the electron.

Inserting Eqs.(\ref{Psi0}) and (\ref{U0}) into
Eq.(\ref{eq:nonloc}) and taking into account that
$\varepsilon_h + \varepsilon_e = E$, we obtain that the
cancellation of the zeroth-order terms in that equation
requires $\lambda   =  \frac{2 \hbar^2 \kappa_h}{m_h}$ and
$\lambda \beta^2  =  \frac{2 \hbar^2 \kappa_e}{m_e}$, from
which we find
\begin{equation}
\left\{
\begin{array}{lcl}
\varepsilon_h &=& \frac{m_h}{M}\, E,\\ \\ \varepsilon_e
&=& \frac{\beta^4 m_e}{M}\, E,\\ \\ \lambda &=& 2\hbar
\sqrt{\frac{2|E|}{M}},
\end{array}
\right. \label{consist}
\end{equation}
which agrees with Eq.(\ref{ratio}) obtained above. Also,
the calculation of the action in the zeroth-order
approximation,
\[
S_0 = \frac{1}{2} \int\!\!dx \left[U_h^2(x) +
U_e^2(x)\right],
\]
gives Eq.(\ref{S0}).

Though the calculation of the first-order correction to
the wave function is, in general, difficult, the
first-order correction to the action can be expressed
through the unperturbed wave functions of the electron and
hole:
\begin{equation}
S_1(r) =  \lambda \int\!\!dx \psi^2_h(x) \left[
\int\!\!dx' \psi^2_e(x') V(x - x' + r) + \beta \lambda
\psi^2_e(x - r)\right],
 \label{S1}
\end{equation}
where $r = x_e - x_h$ is the average electron-hole
separation. The somewhat lengthy derivation of this result
is given in Appendix \ref{appendixB}. The first term in
the square brackets is the correction due to the
electron-hole interaction, while the second term describes
the interaction of the hole with the bump, localizing the
electron and {\em vice versa} (see Eq.(\ref{U0})).

The optimal electron-hole distance $r_\ast$ is found by
minimizing $S_1$ with respect to $r$:
\begin{equation}
\left.\frac{dS_1}{dr}\right|_{r=r_\ast}\!\! = 0.
\label{condr}
\end{equation}
From the form of Eq.(\ref{S1}) it is clear that the latter
condition is just the balance of average forces acting on
the hole: the attraction from the electron and repulsion
from the disorder potential localizing the electron. Thus
the typical electron-hole state of large negative energy
can be considered as a kind of ``molecule'', in which the
disorder fluctuations binding the electron and the hole
play the role of ``nuclei'' with, respectively, positive
and negative charge.

The result Eq.(\ref{S1}) can be cast into a more
transparent form using the fact that for the Coulomb
interaction between the electron and hole to be small
compared to their interaction with the disorder
fluctuations, the optimal electron-hole separation
$r_\ast$ should be large compared to the spatial extent of
the disorder fluctuations: $e^{-\kappa_h r_\ast},
e^{-\kappa_e r_\ast} \ll 1$. We furthermore assume, that
$m_h > m_e \beta^4$ and that $e^{-(\kappa_h - \kappa_e)
r_\ast} \ll 1$, {\em i.e.}, the hole is sufficiently more
localized than the electron, which holds unless $m_h$ is
very close to $m_e \beta^4$. Then Eq.(\ref{S1}) becomes
\begin{equation}
\frac{S_1(r)}{\lambda} \approx - \frac{e^2}{\epsilon r} +
\beta \lambda \psi_e^2(r). \label{S1(r)}
\end{equation}
For the optimal distance $r_\ast$, at which $S_1$ has its
minimum, we obtain
\begin{equation}
\kappa_e r_\ast e^{-\kappa_e r_\ast} =
\sqrt{\frac{e^2}{4\epsilon \beta \lambda}}, \label{rast}
\end{equation}
where $\lambda$ is given by Eq.(\ref{consist}). The
dimensionless parameter on the right-hand side is, essentially,
the square root of the ratio of the Coulomb energy to the
total energy $E$, which, by assumption, is small (and thus
$\kappa_e r_\ast$ is logarithmically large).

Furthermore, the second term in Eq.(\ref{S1(r)}) (due to
repulsion of the hole from the electron optimal
fluctuation) is small compared to the first term (due to
the electron-hole interaction):
\[
\frac{\beta\lambda\psi_e^2(r_\ast)}{e^2/(\epsilon r_\ast)}
\approx \frac{1}{2\kappa_e r_\ast}.
\]
Therefore,
\[
S_1 \approx - \lambda \, \frac{e^2}{\epsilon r_\ast} =
\frac{dS_0}{dE} \,\frac{e^2}{\epsilon r_\ast},
\]
where in the last step we used
\begin{equation}
\lambda = - \frac{dS_0}{dE}, \label{dS0}
\end{equation}
as follows from Eqs.(\ref{S0}) and (\ref{consist}). We
thus see that, to the lowest order, the effect of the
correction to the action $S_1$ is to replace $S_0(E)$ by
$S_0(E + \frac{e^2}{\epsilon r_\ast})$:
\[
F(E) \propto \exp\left\{-\frac{4\sqrt{2}\hbar}{3AM^{1/2}}
\left|E + \frac{e^2}{\epsilon
r_\ast}\right|^{\frac{3}{2}}\right\}.
\]
where the electron-hole Coulomb shift depends on $E$, as
according to Eq.(\ref{rast}), $r_\ast \propto |E|^{-1/2}
\ln |E|$.

To end this section about the shape of the optimal
fluctuation we note that the numerical solution of
Eq.(\ref{condr}) for sufficiently small $|E|$ gives
$r_\ast = 0$, which corresponds to the localization of the
electron-hole pair by a symmetric single-well disorder
potential. In that case the repulsion of the electron from
the disorder potential that localizes the hole is
compensated by the Coulomb interaction between the
electron and hole. The transition from the symmetric to
the asymmetric shape of the optimal fluctuation is studied
in detail Sec.~ref{numres}, where the results of the
numerical solution of Eq.(\ref{eq:nonloc}) are discussed.
We shall show that our analytical approach, in fact, gives
a rather accurate description of that transition (see
Fig.~\ref{re}).

\section{\label{optabs}Optical absorption tail in one dimension}

In the previous section we have shown, that for $d=1$ the
electron and hole in the typical state with a large
negative energy are localized relatively far from each
other. This property allowed us to calculate approximately
the weight of the optimal disorder fluctuation for the
electron-hole pair. It also simplifies the calculation of
the pre-exponential factor in the expression for the
absorption rate, which results from the integration over
small deviations from the optimal disorder fluctuation. In
this section we obtain an analytical expression for the
tail of the optical absorption spectrum, which actually
has a wider range of validity than the standard optimal
fluctuation method.

As we have shown in the previous section, the dominant
contribution to the optical absorption at large negative
$E$ comes from the disorder realizations that are close to
the sum of the single-particle optimal fluctuations,
$U_0(x) = U_h(x - x_h) + U_e(x - x_e)$ (see
Eq.(\ref{U0})), where $U_h(x-x_e)$ localizes the hole with
energy $\varepsilon_h$ near $x = x_h$ and $U_e(x-x_e)$
localizes the electron with the energy $\varepsilon_e$
near $x = x_e$. In the previous section we have calculated
the optimal electron-hole separation, for which the weight
of the disorder fluctuation reaches its maximum. In this
section we shall treat $x_h$ and $x_e$, as well as the
single-particle energies $\varepsilon_h$ and
$\varepsilon_e$,  as the collective variables of the
functional integration over disorder. Performing the
Gaussian integration over all the small deviations $\delta
U(x) = U(x) - U_0(x)$ in Eq.(\ref{F}) that are orthogonal
to the deviations corresponding to the four collective
coordinates, we obtain
\begin{eqnarray}
F(E) &\simeq& \frac{1}{L}
\int d\varepsilon_h d\varepsilon_e dx_h dx_e
K_h(\varepsilon_h) K_e(\varepsilon_e)
D^2(x_e-x_h)
\nonumber \\
&\times& e^{-\frac{1}{A}
\left(S_h(\varepsilon_h) + S_e(\varepsilon_e)+\delta S\right)}
\delta(\varepsilon_h+\varepsilon_e+\delta \varepsilon-E).
\label{F1}
\end{eqnarray}
Here, $L$ is the chain length, $D$ is the transition
matrix element (see Eq.(\ref{D})), $S_h(\varepsilon_h)$
and $S_e(\varepsilon_e)$ are the single-particle actions,
given by Eq.(\ref{action2}), and $\delta S = \delta S(r)$, where
 $r = x_e - x_h$, is the correction to the action of the
electron-hole pair due to the overlap between the electron and
hole optimal fluctuation:
\[
\delta S(r) = \int\!\!dx U_h(x - x_h) U_e(x - x_e).
\]
Furthermore, $\delta \varepsilon = \delta \varepsilon(r)$ is the
energy correction due to the electron-hole interaction and the
interaction of the electron with hole optimal fluctuation and
{\em vice versa}. The first-order correction, calculated using
the unperturbed electron-hole wave function
Eq.(\ref{Psi0}), is
\begin{eqnarray}
\delta \varepsilon(r)
&=&\!\!\int\!\!dx\left[\psi_h^2(x-x_h) U_e(x-x_e) +
\psi_e^2(x-x_e) U_h(x-x_h) \right]\nonumber\\
\!\!\!&+&\!\!\! \int\!\! dxdx' \psi_h^2(x-x_h) V(x - x')
\psi_e^2(x'-x_e),
\end{eqnarray}
where $r = x_e - x_h$ is the electron-hole separation. In
the same approximation the transition matrix element is
given by
\[
D(r) = \int \!\!dx \psi_h(x-x_h) \psi_e(x-x_e).
\]
Finally, in Eq.(\ref{F1}) we used that for small overlap
between the electron and hole optimal fluctuations the
prefactor, resulting from the Gaussian integration over
$\delta U(x)$, is the product of prefactors for isolated
hole and electron, $K_h(\varepsilon_h) K_e(\varepsilon_e)
$.\cite{instanton} (We note, that in $K_e(\varepsilon_e)$
the disorder strength $A$ has to be substituted by
$\beta^2 A$.)

The integration over the center-of-mass coordinate
\[
R = \frac{m_h x_h + m_e x_e}{m_h + m_e},
\]
which is a zero mode, gives the chain length $L$. Due to
the $\delta$-function in Eq.(\ref{F1}) $\varepsilon_e = E
- \delta \varepsilon - \varepsilon_h$ and the remaining
integration over the hole energy $\varepsilon_h$ can be
performed in the saddle-point approximation. The condition
of the minimum of $S_h(\varepsilon_h) +
S_e(\varepsilon_e)$ gives Eq.(\ref{ratio}), which
determines the saddle-point values of $\varepsilon_h$ and
$\varepsilon_e$. Then Eq.(\ref{F1}) can be written as
follows:
\begin{equation}
        F(E) \simeq \sqrt{\frac{2\pi A}
        {B}}
        \int dr K_h(\varepsilon_h)  K_e(\varepsilon_e)
         D^2(r)
        e^{-\frac{1}{A}
        \left(S_0(E - \delta \varepsilon) + \delta
            S\right)},
        \label{F2}
\end{equation}
where
\[
B = \frac{d^2}{d\varepsilon_h^2}\left[S_h(\varepsilon_h) +
S_e(E - \varepsilon_h)\right] = \frac{\hbar}{\mu} \sqrt
\frac{2M}{{|E|}},
\]
$\mu = \frac{\beta^4 m_e m_h}{m_h + \beta^4m_e}$, and
$S_0(E)$ is given by Eq.(\ref{S0}).

\begin{figure}
    \includegraphics[width=7cm]{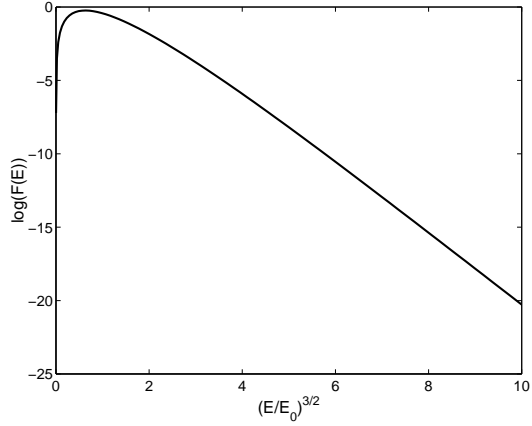}
    \caption{\label{FE}The energy dependence of the logarithm of the
    absorption rate.}
\end{figure}

Expanding $S_0(E - \delta \varepsilon(r)) \approx S_0(E) -
\frac{dS_0}{dE} \delta \varepsilon(r)$ and using
Eq.(\ref{dS0}) and the relation between the
single-particle optimal fluctuations and the wave
functions: $U_h(x - x_h) = - \lambda \psi_h^2(x-x_h)$ and
$U_e(x-x_e) =  \lambda \beta \psi_e^2(x-x_e)$, we find
that the $r$-dependent part of the action coincides with
the first-order correction to the action of the optimal
fluctuation Eq.(\ref{S1}):
\[
\delta S - \frac{dS_0}{dE} \delta \varepsilon = S_1(r).
\]
Thus, we can write Eq.(\ref{F2}) in the form
\begin{equation}
         F(E) \simeq \sqrt{\frac{2\pi A}
        {B}}
        \int dr K_h(\varepsilon_h)  K_e(\varepsilon_e)
         D^2(r)
        e^{-\frac{1}{A}
        \left(S_0(E) + S_1(r)\right)}.
        \label{F3}
\end{equation}
The typical energy dependence of the absorption rate
Eq.(\ref{F3}) is shown in Fig.\ref{FE}, where we plot
$\log F$ as a function of $\left(E/E_0\right)^{3/2}$, with
$E_0<0$ being the exciton binding energy. One can see,
that the energy-dependence of $\log F$ very quickly
becomes linear for energies below $E_0$. The linear
dependence reflects the relative weakness of the Coulomb
interaction between the electron and hole (see
Eq.(\ref{S0})), which was the main assumption of this
analytical calculation.

In the case when the integral over the electron-hole
separation $r$ comes from a small vicinity of the
saddle-point $r_\ast$, determined by Eq.(\ref{condr}), we
find that the saddle-point action coincides with the
action for the optimal fluctuation, $S_0 + S_1(r_\ast)$,
obtained in the Sec.~\ref{optfluc}. The result of the
saddle-point integration over $r$ is
\[
F(E) \simeq \frac{2\pi A}{\sqrt{B
\frac{d^2S}{dr_{\ast}^2}}} \rho_h(\varepsilon_h)
\rho_e(\varepsilon_e) D^2(r_\ast)
e^{-\frac{S_1(r_\ast)}{A}},
\]
where
\[
\frac{d^2S}{dr_{\ast}^2} \approx \frac{2\lambda
e^2}{\epsilon r_{\ast}^3} \left(\kappa_e r_\ast - 1\right)
\]
and  $\varepsilon_h$ and $\varepsilon_e$ are the
unperturbed single-particle energies, given by
Eq.(\ref{consist}) (we have used that the factors
$K_h(\varepsilon_h)$ and $K_e(\varepsilon_e)$ are slow
functions of the energies).

Equation (\ref{F3}) also applies when, instead of one
optimal fluctuation, one finds an entire manifold of
disorder realizations that significantly contribute to the
optical absorption. In particular, for a non-interacting
electron-hole pair the minimum of the action $S_1$ is
reached at the largest possible electron-hole separation,
since for $V(x_1 - x_2) = 0$ nothing can counteract the
``repulsion'' between the two disorder fluctuations,
described by the second term in Eq.(\ref{S1}). However,
these electron-hole states clearly do not contribute to
the optical absorption as the transition matrix element
$D$ vanishes at infinite $r$. Since the repulsion decays
exponentially with the electron-hole separation, the
action $S_1$ is a very weak function of $r$, as soon as
the latter exceeds the spatial extent of the electron and
hole states, $\kappa_h^{-1}$ and $\kappa_e^{-1}$. The
contribution of the electron-hole pairs with large $r$ is
then suppressed not by the weight of such fluctuations,
but by the smallness of the transition matrix element $D$,
which decays exponentially with $r$. Thus, when the
electron-hole interaction is absent or relatively weak, to
calculate the absorption rate, one has to sum the
contributions of many disorder fluctuations with different
electron-hole separations, which can be accomplished using
Eq.(\ref{F3}).

The integration in Eq.(\ref{F3}) for non-interacting
electron and hole becomes particularly simple in the case
when the hole is localized stronger than the electron, or
more precisely, when $e^{-(\kappa_h - \kappa_e)r} \ll 1$
for relevant electron-hole separations $r$, which we
already have used in the Sec.~\ref{optfluc}. Then the
action $S_1$ for the non-interacting case is \[ S_1(r) =
\lambda^2 \beta\int\!\!dx \psi_h^2(x) \psi_e^2(x+r)
\approx \lambda^2 \beta \psi_e^2(r)
\]
and the transition matrix element is given by
\[
D(r) \approx \frac{\pi}{\sqrt{2\kappa_h}} \psi_e(r)
\]
and
\[
\int\!\!dr D(r)^2 e^{-\frac{S_1(r)}{A}} =
\frac{\pi^2}{4\kappa_h}
\int_{-\infty}^{+\infty}\!\!\frac{dz}{\cosh^2
z}\,e^{-\frac{\zeta}{\cosh^2 z}} \approx \frac{\pi^2 \beta
M A}{32 \mu |E|^2},
\]
where we assumed that
\begin{equation}
\zeta = \frac{\lambda^2\beta\kappa_e}{2A} = \frac{3 m_e
\beta^3}{M} \frac{S_0}{A} \gg 1. \label{zeta}
\end{equation}
Since for the applicability of the optimal fluctuation
method, anyhow, the action $S_0$ has to be much larger
than $A$, Eq.(\ref{zeta}) holds, unless  $\frac{3 m_e
\beta^3}{M}$ is very small.

Thus, finally, for non-interacting  electron-hole pairs
the absorption rate is given by
\[
F(E) \simeq \frac{1}{\beta} \left[\frac{\pi \mu}{2A\hbar}
\sqrt{\frac{|E|}{2M}}\right]^{\frac{1}{2}}
\exp\left\{-\frac{4\hbar}{3A}
\sqrt{\frac{2|E|^3}{M}}\right\}.
\]

\section{\label{numres}Numerical results}

The analytical results of sections \ref{optfluc} and
\ref{optabs} for the optimal fluctuation and the optical
absorption spectrum were obtained by a perturbative
treatment of the Coulomb interaction between the electron
and hole. When this interaction is of the same order as
the magnitude of the disorder potential the optimal
fluctuation has to be solved numerically. In this section
we present our numerical results for the optimal
fluctuation in one dimension.

Instead of solving Eq.(\ref{eq:nonloc}) directly, we
perform the numerical calculations for a tight-binding
model defined on a lattice.
%which only reduces to the
%continuum model, defined by Eqs.(\ref{Sch}) and
%(\ref{whitenoise}), in certain limit.
The discrete version of the Schr\"odinger equation
(\ref{Sch}) reads:
\begin{equation}
    - t_1(\psi_{n-1,m} +\psi_{n+1,m}) - t_2(\psi_{n,m+1} + \psi_{n,m-1})
 + V_{nm}\psi_{nm} +(U_n-\beta U_m)\psi_{nm} = {\cal
 E}\psi_{nm},
    \label{eq:disc}
\end{equation}
where $t_1$ and $t_2$ are, respectively, the hole and
electron hopping amplitudes, the indices $n,m = 1,2,
\ldots,L$ denote the sites of the one-dimensional lattice
with the lattice constant $a$ and periodic boundary
conditions, $U_n$ is the disorder potential:
\[
\langle U_n \rangle = 0, \;\;\;\;\langle U_n U_m \rangle =
\frac{A}{a} \,\delta_{n,m},
\]
and $V_{nm}$ is the regularized Coulomb interaction
\begin{equation}
V_{nm} = - g_0 \left( \frac{1}{|n-m|+\delta_{n,m}} +
\frac{1}{N - |n - m|+\delta_{|n-m|,N}} \right).
\label{regcoul} \label{Vnm}
\end{equation}
Here, $g_0 = \frac{e^2}{\epsilon a}$ and the second term
in the brackets is added to satisfy the periodic boundary
conditions.

%Continuum limit
The discrete equation (\ref{eq:disc}) reduces to the
Schr\"odinger equation (\ref{Sch}) in the continuum limit,
when all relevant electron and hole states have small wave
vectors, $k_1a, k_2 a \ll 1$. In that case, the dispersion
of the free hole dispersion is
\[
\varepsilon_h(k) = - 2 t_1 \cos ka \approx - 2 t_1 + t_1
(ka)^2,
\]
so that $m_h = \frac{\hbar^2}{2 t_1 a^2}$ and, similarly,
for the electron we have $m_e = \frac{\hbar^2}{2 t_2
a^2}$. In this calculation we put $a = t_1 = t_2 = 1$.
Then the continuum limit is reached for small values of
the coupling constant $g_0$ and the energy $E$, counted
from the bottom of the band:
\[
E = {\cal E} + 2 (t_1 + t_2) = {\cal E} + 4 \ll 1.
\]

The discrete analog of the action (\ref{Slam1}) is
\begin{equation}
S_\lambda[\psi,U] = \frac{1}{2}\sum_n U_n^2
+\lambda\left(E[\psi,U] - {\cal E} \right) \label{Slam2},
\end{equation}
where
\begin{equation}
      E[U,\psi] = \frac{-2\sum_{nm} \left[\psi_{nm}( t_1\psi_{n+1,m} +
      t_2 \psi_{n,m+1}) + V_{nm}\psi_{nm}^2 + (U_n-\beta U_m)\psi_{nm}^2\right]}
  {\sum_{nm}\psi_{nm}^2} .
\end{equation}
The denominator in the last equation takes care of the
wave function normalization. It is readily seen that
varying $S_\lambda$ with respect to the wave function
$\psi_{nm}$ (that can be chosen real) yields the discrete
Schr\"{o}dinger equation (\ref{eq:disc}), while the
minimization of $S_\lambda$ with respect to the disorder
potential $U_n$ gives a relation between the optimal
fluctuation and $\psi_{nm}$,
\begin{equation}
      U_n = -\frac{\lambda}{\cal N} \sum_m (\psi_{nm}^2-\beta\psi_{mn}^2)
      \label{eq:discopt} ,
\end{equation}
where we have used the notation
\begin{equation}
   {\cal N} = \sum_{nm}\psi_{nm}^2.
\end{equation}

We find the optimal fluctuation $U_n$ and the
corresponding wave function $\psi_{nm}$ by minimizing the
functional
\begin{eqnarray}
     {\cal A}_\lambda[\psi] = \!\!&-&\!\!\!\frac{2}{\cal N} \sum_{nm} \psi_{nm}(t_1\psi_{n+1,m}
     + t_2\psi_{n,m+1}) +\frac{1}{\cal N}\sum_{nm} V_{nm}\psi_{nm}^2
     \nonumber\\
     \!\!\!&-&\!\!\!
     \frac{\lambda}{2 {\cal N}^2}\sum_n\left(\sum_m(\psi_{nm}^2 -
       \beta\psi_{mn}^2) \right)^2,
\end{eqnarray}
which is the two-particle analog of Eq.(\ref{A}). While
$S_\lambda$ depends on both $\psi_{nm}$ and $U_n$, ${\cal
A}_\lambda$ is a functional of the wave function only. One
can easily check, that the condition $\frac{\delta {\cal
A} _\lambda}{\delta \psi_{nm}} = 0$ is equivalent to
Eq.(\ref{eq:disc}) with the disorder potential $U_n$ given
by Eq.(\ref{eq:discopt}). The minimization of ${\cal A}
_\lambda$ with respect to $\psi_{nm}$ was carried out
numerically, using the steepest descent algorithm for $L =
50$. The energy of the electron-hole pair and other
quantities of interest, {\em e.g.}, the optimal
fluctuation $U_n$, its weight, and the corresponding
electron-hole wave function, are first obtained as
functions of the Lagrangian multiplier $\lambda$. Then we
eliminate $\lambda$ by replotting these quantities as
functions of the energy $E$.

\begin{figure}
    \includegraphics[width=7cm]{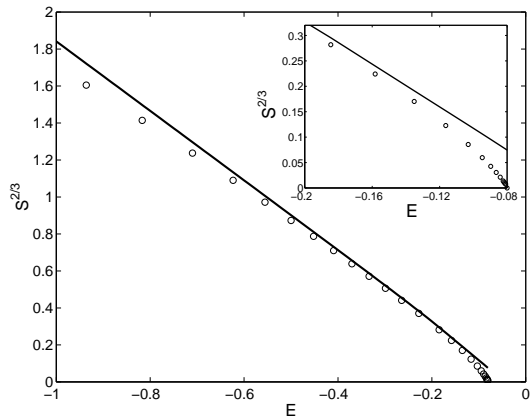}
    \caption{\label{fig:weight}The action $S$ of the optimal fluctuation to the power
    $2/3$ plotted as a function of the electron-hole energy $E$ for
    $g_0 = 0.2$, $\beta = 0.5$ and the chain length $L = 50$.
    The open circles are results of the numerical calculation and the
    solid curve was obtained analytically (see
    explanations in the text).}
\end{figure}

We first consider the energy dependence of the weight of
the optimal fluctuation, which, essentially, determines
the energy dependence of the optical absorption rate. This
weight is given by $e^{-\frac{S}{A}}$, where $S =
\frac{1}{2} \sum_n U_n^2$ is the action of the optimal
fluctuation [cf. Eq.(\ref{defaction})]. Motivated by the
dominant $|E|^{3/2}$ behavior of $S$ [cf. Eq.(\ref{S0})],
we plot in Fig.~\ref{fig:weight} $S^{2/3}$ as a function
of the energy $E$ for $g_0 = 0.2$ and $\beta = 0.5$. The
open circles are obtained by the numerical procedure
described above, while the solid line is the result of our
approximate analytical calculation of the action: $S =
S_0+ S_1$, where $S_0$ and $S_1$ are given by Eqs.
(\ref{S0}) and (\ref{S1}), respectively. Clearly, apart
from a small energy interval near the exciton binding
energy in the absence of disorder, $E_0 \approx -0.08$,
the energy dependence of $S^{2/3}$ is indeed close to
linear and the agreement between our numerical and
analytical results is good.

For energies close to $E_0$, the assumption that disorder
dominates the Coulomb interaction, used in our analytical
approach, breaks down, which explains the deviations of
the numerical data from the $S \propto |E|^{3/2}$ law and
from the analytical curve. On the other hand, the
deviations found at relatively large energies $|E| \sim 1$
are due to the break down of the continuum approximation,
resulting from the fact that the hole becomes localized on
a single lattice site.

\begin{figure*}
    \includegraphics[width=14cm]{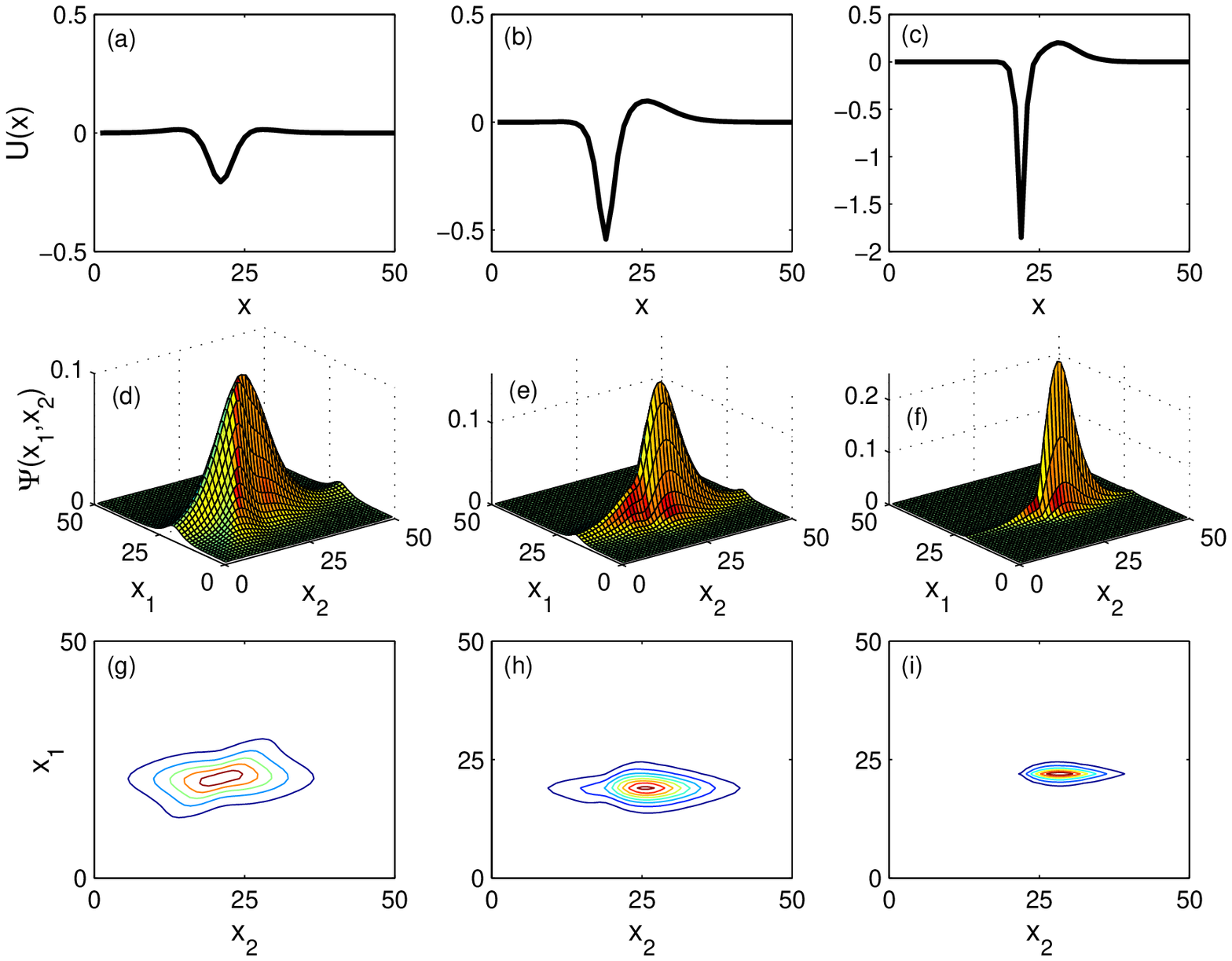}
    \caption{\label{fig:optflu}The shape
     of the numerically obtained optimal fluctuation for $E = -0.13$ (a), $E = - 0.3$ (b),
     and $E = - 0.94$ (c). Panels (d - f) show
     the corresponding electron-hole wave function
     $\psi_{nm}$ and panels (g - i) show the
     contour plot of the wave function.  All plots correspond to $g_0 = 0.2$,
    $\beta=0.5$, and $L = 50$. The coordinates $x_1 = na$ and
    $x_2 = ma$, where $n,m = 1,\ldots,L$ describe, respectively,
    the hole and electron positions in the chain in units of $a=1$. Note, that
    as $E$ decreases, the shape of the optimal fluctuation
    undergoes a transition from a single-well to a
    ``dip-bump'' structure.}
\end{figure*}

These changes in the energy-dependence of $S$ reflect
changes in the shape of the optimal fluctuation and the
corresponding electron-hole wave function. In
Fig.~\ref{fig:optflu} we plot $U_n$ and $\psi_{nm}$,
calculated numerically, for three different values of the
energy: $-0.14$, $-0.30$, and $-0.94$. The first value $E
= -0.14$ is rather close to the exciton binding energy
$E_0$. In that case, the optimal fluctuation, shown in
Fig.~\ref{fig:optflu}a, is rather shallow and it is
symmetric around $x_0$, where $x_0$ is the position of the
minimum of this fluctuation. This symmetry implies that no
separation exists between the average electron and hole
positions. Such an optimal fluctuation was discussed in
Ref.~\onlinecite{EfrosRaikh} for the situation where the
Coulomb interaction dominates the disorder and the spatial
extent of the optimal fluctuation $l_{opt}$ is much larger
than the exciton radius $r_{ex}$, leading to decoupling of
the center-of-mass and the relative motion. That limit is
rather difficult to simulate numerically within our
discrete model, as the requirement to maintain the
validity of the continuum approximation then leads to $1
\ll r_{ex} \ll l_{opt} \ll L$, thus forcing us to consider
a very large lattice size $L$. From Fig.~\ref{fig:optflu}
one observes that for $E = -0.14$, $l_{opt}$ is comparable
to the exciton radius $r_{ex}\sim 5$. Still, one can see
from Figs.~\ref{fig:optflu} d and g, that the
electron-hole wave function, apart from the delocalization
along the electron coordinate $x_2$, also shows a strong
delocalization along the line $x_1 = x_2$, which
corresponds to the center-of-mass motion of the exciton.
For the second value of the energy, $E = -0.3$, the
optimal fluctuation has the asymmetric ``dip-bump'' shape,
which corresponds to the localization of the electron and
hole on different sites of the chain (see
Fig.~\ref{fig:optflu}b). Finally, at $E = -0.94$, the hole
is practically localized on one chain site, in which case
the discrete model should not be used to simulate the
continuum one (see Fig.~\ref{fig:optflu}c). From Figs.
\ref{fig:optflu} e, f,  h, and i, one can see that for $E
= -0.3$ and $E = -0.94$ the electron-hole wave function
$\psi_{nm}$ is mostly delocalized along the electron
coordinate $x_2$ and it has no delocalization along the
center-of-mass direction, $x_1 = x_2$.

\begin{figure}
\includegraphics[width=7cm]{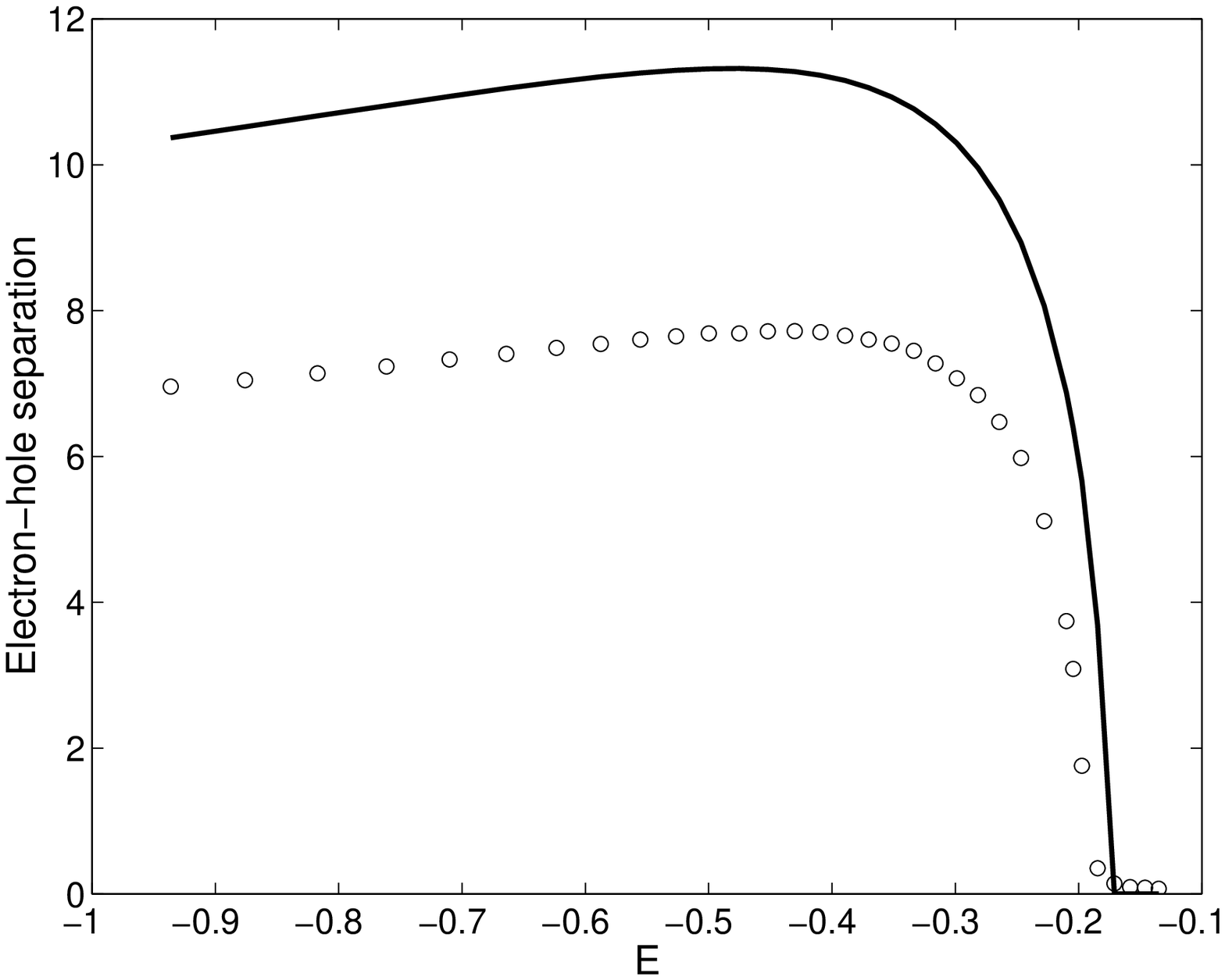}
\caption{\label{re} Numerical (circles) and analytical
(solid line) energy-dependence of the average
electron-hole separation calculated for $g_0 = 0.2$,
$\beta = 0.5$, and $L = 50$.}
\end{figure}

The cross-over from the symmetric to the asymmetric shape
of the optimal fluctuation can be most clearly seen from
the energy dependence of the average electron-hole
separation, defined by
\[
r = \sum_{nm}(m-n) \psi_{nm}^2.
\]
The results of the numerical calculation of $r$ as a
function of the energy $E$ are shown in Fig.~\ref{re} by
circles. In that figure we also plot the optimal
electron-hole distance $r_\ast$, obtained by our
approximate analytical approach of Sec.~\ref{optfluc}
[Eq.(\ref{condr})]. The minimization of the correction to
the action $S_1$  was performed numerically, as the
approximation that was used to obtain Eq.(\ref{rast}) is
too crude to describe the changes in the shape of the
optimal fluctuation. Figure~\ref{re} shows a very fast
transition from the optimal fluctuation with zero average
electron-hole separation to one with finite separation.
One can also see that the approximate analytical approach
provides a good qualitative description of this cross-over
(in particular, the onset of the cross-over and the shape
of the $r(E)$ curve), but it gives a somewhat larger value
of the electron-hole separation at low energies.

\begin{figure}
    \includegraphics[width=7cm]{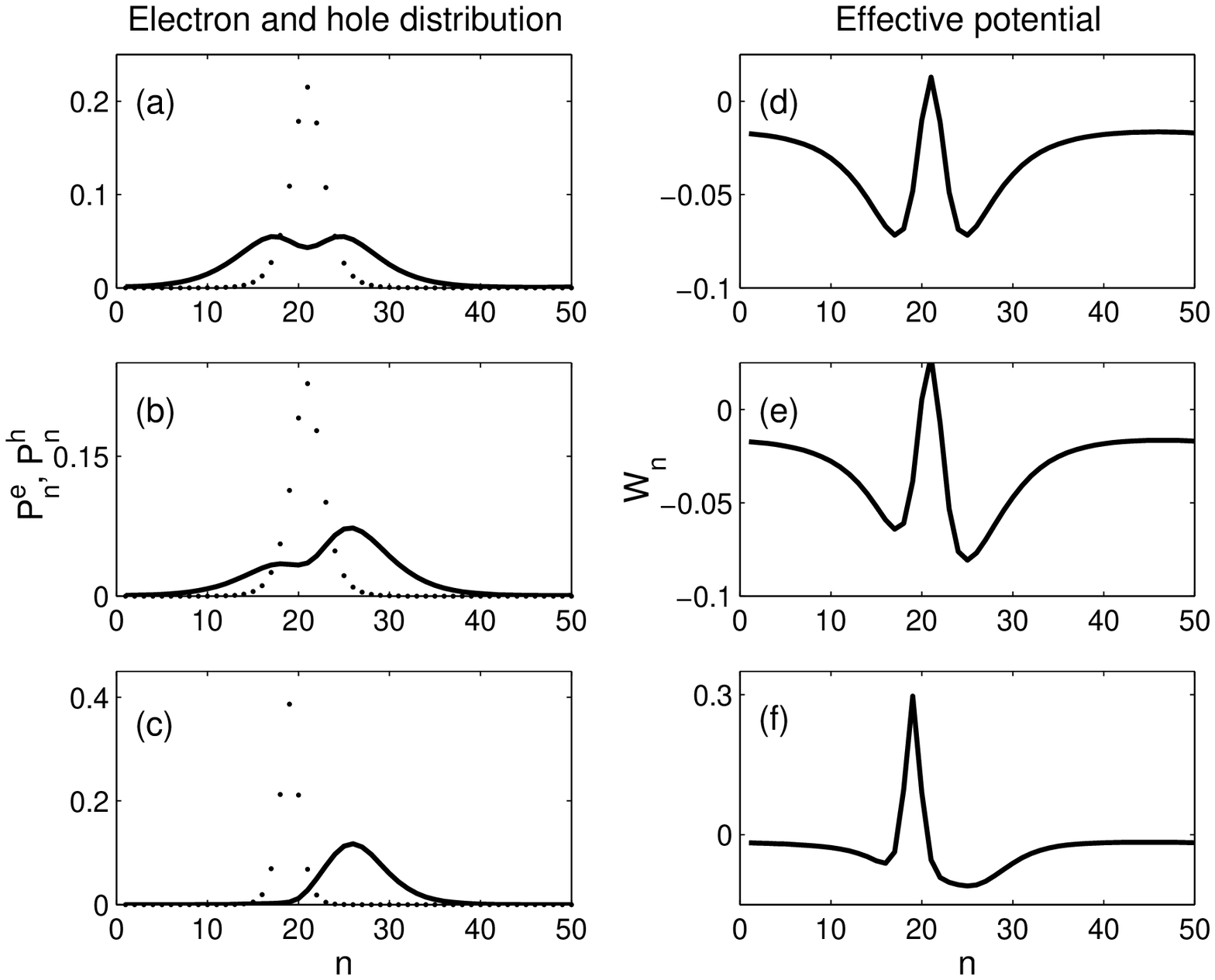}
    \caption{\label{dist}
The left-hand side of the figure shows the electron (solid
line) and the hole (points) distributions in the typical
electron-hole state for $E = -0.18$ (a), $E = - 0.20$ (b),
and $E = - 0.50$ (c). On the right-hand side (panels (d) -
(f)) the corresponding effective potentials acting on the
electron are plotted. All plots correspond to $g_0 = 0.2$,
$\beta=0.5$, and $L = 50$.}
\end{figure}

To clarify the nature of this cross-over, we plot in
Fig.~\ref{dist}(a-c) the coordinate distributions of the
electron (solid line) and hole (dotted line),
\[
\left\{
\begin{array}{rcl}
P^h_n & = & \sum_m\psi^2_{nm},\\ \\ P^e_m & = &
\sum_n\psi^2_{nm},
\end{array}
\right.
\]
for $g_0 = 0.2$, $\beta = 0.5$, and three different values
of energy $E$. In addition, Figs.~\ref{dist}(d-f) show the
corresponding effective Hartree potential acting on the
electron
\[
W_m = -\beta U_m + \sum_n P^e_n V_{nm}.
\]
At $E = -0.18$, just above the cross-over, the electron
wave function has two peaks, in accordance with the
double-well structure of the effective Hartree potential
(see Figs.~\ref{dist} a and d). The peak separating the
two potential wells is the disorder potential that
localizes the hole and repels the electron. As the energy
$E$ decreases, the height of the peak grows, which
suppresses the electron tunneling between the two wells.
For large separations between the wells, the weight of the
symmetric optimal fluctuation, in which the electron is
delocalized over the two wells, is lower than the weight
of the single-well optimal fluctuation with the same
electron energy. For $E = -0.20$ in the cross-over region
the two wells become unequal [see Figs. \ref{dist}b and
\ref{dist}e], and at $E = -0.50$, well below the
cross-over region, the electron is predominantly located
in a single well [see Figs. \ref{dist}c and \ref{dist}f].

\section{\label{conclude}Conclusions}

In this paper we studied theoretically the photoexcitation
of electron-hole pairs in disordered one-dimensional
semiconductors. Using the optimal fluctuation method, we
calculated the low-energy tail of the optimal absorption
spectrum in these systems. We were in particular
interested in the effects of the Coulomb interaction
between the electron and hole on the energy dependence of
the absorption spectrum. We want to point out, however,
that the calculation of the absorption rate is a
nontrivial problem even for non-interacting electron and
hole. In particular, it cannot be reduced to a
single-particle calculation, since, on the one hand, the
electron and hole move in a common disorder potential,
and, on the other hand, the effect of disorder on the
electron is different from the effect of disorder on the
hole.

We found that, as the photon energy decreases, the shape
of the optimal fluctuation undergoes a cross-over. Close
to $E_0$ (the exciton binding energy in the absence of
disorder), the Coulomb energy dominates disorder and the
optimal fluctuation has a symmetric shape. It reflects the
fact that at those energies the exciton is not entirely
destroyed by disorder. That limit was considered
previously in Ref.~\onlinecite{EfrosRaikh}. In the
opposite limit, when the disorder dominates the Coulomb
interaction, the optimal fluctuation has two parts: a
``dip'' that localizes the positively charged hole and a
``bump'' that localizes the negatively charged electron.
This cross-over to an asymmetric optimal fluctuation is
characteristic for motion in one dimension - it does not
occur in two and three dimensions. This cross-over can, in
principle, be observed experimentally, as the absorption
rate has a different energy-dependence above and below the
cross-over energy (see inset in Fig.~\ref{fig:weight}).

At photon energies well below the exciton energy, when the
Coulomb interaction can be treated as a perturbation, we
obtained an analytical expression for the optical
conductivity of a disordered one-dimensional
semiconductor. We showed that in some cases one has to go
beyond the standard optimal fluctuation method and
consider an entire manifold of optimal fluctuations,
corresponding to different electron-hole separations. We
also performed numerical calculations of the optimal
fluctuation for a discrete model that allowed us to study
the whole region of photon energies, in which the optimal
fluctuation method is applicable ({\em i.e.}, for $E <
E_0$, for which the action $S$ is much larger than $A$).
In the continuum limit we found a good agreement between
our numerical and analytical results.

\begin{acknowledgments}
This work is financially supported by the Stichting voor
Fundamenteel Onderzoek der Materie, FOM.
\end{acknowledgments}

\appendix

\section{Saddle-point integration}

The integration over disorder realizations close to the
optimal one for a particle in a one-dimensional
white-noise potential was performed by a number of
different methods.\cite{LGP} Here we perform this
integration by reducing it to a path-integration over the
trajectories of a particle moving in the double-well
potential. A more detailed discussion of this method for a
similar problem is contained in Ref.~\cite{mj}.

To find the average density of states, we have to perform
the functional integration
\begin{equation}
\langle \rho(\varepsilon) \rangle = \frac{1}{L} \int {\cal
D} U \left[\sum_{\alpha} \delta(\varepsilon_\alpha[U] -
\varepsilon)\right] e^{-\frac{1}{2A} \int_{-L/2}^{+L/2}dx
U^2}. \label{rhoU}
\end{equation}
In this appendix we put $\frac{\hbar^2}{2m} = 1$, so that
Eq.(\ref{eq:schr1}) becomes
\begin{equation}
\left(-\frac{d^2}{dx^2} + U(x)\right)\psi_\alpha(x) =
\varepsilon_\alpha \psi_\alpha(x), \label{sch1}
\end{equation}
and $\varepsilon = -\kappa^2$. We consider Eq.(\ref{sch1})
on the interval $x \in \left
[-\frac{L}{2},+\frac{L}{2}\right]$.

The correspondence between the disorder average and the
quantum-mechanical average is established through the
functional substitution $U(x) \rightarrow y(x)$:
\begin{equation}
U(x) = -\frac{dy}{dx} + y^2 + \varepsilon. \label{subs}
\end{equation}
Then the action for the disorder realization $U(x)$ can
written in the form
\begin{eqnarray}
S &\!\!\!=\!\!\!&
\frac{1}{2}\int_{-L/2}^{+L/2}\!\!dxU^2(x)\nonumber\\
  &\!\!\!=\!\!\!& \frac{1}{2} \int_{-L/2}^{+L/2}\!\!dx
  \left[ \left(\frac{dy}{dx}\right)^2 +
  \left(y^2-\kappa^2\right)^2\right] + \left(\kappa^2 y -
  \frac{y^3}{3}\right)\bigg|_{-L/2}^{+L/2}.
  \label{cor}
\end{eqnarray}
The integral in the second line of Eq.(\ref{cor}) can be
considered as the Euclidean action of a particle with the
coordinate $y$ moving in the double-well potential $W(y) =
\frac{1}{2}\left(y^2-\kappa^2\right)^2$. Thus, the
functional integration over disorder realizations $U(x)$
can be represented as the path-integral over trajectories
of the particle $y(x)$ in the imaginary time $x$.

The regularized form of the functional substitution
Eq.(\ref{subs}), obtained by discretizing the imaginary
time $x_n = -\frac{L}{2} + a n$, where $a = \frac{L}{N}$
and $n = 0,1,\ldots,N$, is
\[
\frac{(y_n - y_{n-1})}{a} = \frac{y_{n-1}^2 + \varepsilon
- U_n}{1 - a y_{n-1}}.
\]
Here $y_n = y(x_n)$, $U_n = U(x_n)$, and terms of order
$a^2$ are neglected. The last equation is a one-to-one
correspondence between $U(x)$ and $y(x)$ if the value of
$y$ at the left boundary of the interval of $x$ is kept
fixed:
\[
y_0 = y(-L/2) = y_L,
\]
where $y_L$ is an arbitrary number. The Jacobian of this
functional substitution is given by
\[
J = \frac{{\cal D} U}{{\cal D} y} =
\exp\left[-\int_{-L/2}^{+L/2} dx y(x)\right].
\]

Furthermore, if, for a given $U(x)$, $\varepsilon =
\varepsilon_\alpha$ is the energy eigenvalue, then $y(x)$,
defined by Eq.(\ref{subs}), is related to the
eigenfunction $\psi_\alpha(x)$ by
\[
y(x) = - \frac{\psi'_\alpha(x)}{\psi_\alpha(x)}.
\]
In that case $y(x)$ also satisfies a condition at the
right boundary $+\frac{L}{2}$:
\[
y_N = y(+L/2) = y_R.
\]
Therefore, the density of the single-particle states can
be written in the form
\[
\sum_{\alpha} \delta \left(\varepsilon -
\varepsilon_\alpha\right) = \left| \frac{\partial
y(L/2)}{\partial \varepsilon}\right| \delta\left(y(L/2) -
y_R\right),
\]
where the derivative $\frac{\partial y}{\partial
\varepsilon}$ at $x = +L/2$ is calculated at fixed
$y(-L/2)$ and $U(x)$. Differentiating Eq.(\ref{subs}) with
respect to $\varepsilon$, we obtain
\[
\left(- \frac{\partial}{\partial x} + 2 y(x)\right)
\frac{\partial y}{\partial \varepsilon} = -1,
\]
the solution of which is
\[
\frac{\partial y(L/2)}{\partial \varepsilon} = z(L/2)
\int_{-L/2}^{x} \frac{dx'}{z(x')},
\]
where
\[
z(x) = \exp\left[2\int_{-L/2}^{x} dx' y(x')\right].
\]

Thus, Eq.(\ref{rhoU}) can be written in the path-integral
form
\begin{equation}
\langle \rho(\varepsilon) \rangle = \frac{1}{L}\int\!\!
{\cal D}'y G[y] e^{-\frac{S[y]}{A}},
\end{equation}
where $S[y]$ is given by Eq.(\ref{cor}),
\[
G[y] = \sqrt{z(L/2)} \int_{-L/2}^{+L/2} \frac{dx}{z(x)},
\]
and the measure $\int\!{\cal D}'y \propto \sum_{n=1}^{N-1}
dy_n$ does not include the integrations over the values of
$y$ on the left and right boundaries.

According to Eq.(\ref{subs}), the optimal fluctuation
\begin{equation}
{\bar U}(x) = - \frac{2\kappa^2}{\cosh^2(\kappa x)}
\label{Uopt}
\end{equation}
(see Eq.(\ref{eq:optU})) corresponds then to the instanton
trajectory
\begin{equation}
{\bar y}(x) = \kappa \tanh(\kappa x), \label{yinst}
\end{equation}
which describes the tunneling from the bottom of the left
well, $y = -\kappa$, to the bottom of the right well, $y =
+ \kappa$ and is the saddle-point trajectory of the
Euclidean action (\ref{cor}). Then the integration over
disorder configurations close to the optimal fluctuation
${\bar U}$ reduces to the saddle-point integration of the
imaginary-time trajectories close to the instanton
trajectory ${\bar y}$. The result of this integration is
\[
\langle \rho(\varepsilon) \rangle = \frac{1}{L} G[{\bar
y}] {\cal M}^{-\frac{1}{2}} e^{-\frac{S[{\bar y}]}{A}},
\]
where
\[
{\cal M} = \det\left[-\frac{d^2}{dx^2} + \frac{1}{2}
W''({\bar y}) \right].
\]

In the limit large $L$, we obtain $S[{\bar y}] \simeq
\frac{4}{3} \kappa^3$, $ G[{\bar y}] \simeq
\frac{e^{\kappa L}}{2\kappa}$, and the functional
determinant ${\cal M}$ can be found using the asymptotic
behavior of the instanton solution Eq.(\ref{yinst}) at
large distances:\cite{instanton}
\[
{\cal M}^{-\frac{1}{2}} = L \frac{8\kappa^3}{\pi A}
e^{-\kappa L},
\]
where the factor $L$ comes from the integration over the
instanton zero mode, which corresponds to the fact that
the optimal fluctuation can be located at any place within
the chain of size $L$. Combining these results, we obtain
\[
\langle \rho(\varepsilon) \rangle =
\frac{4|\varepsilon|}{\pi A} e^{-\frac{8
|\varepsilon|^{3/2}}{3A}}
\]
or, in usual units,
\begin{equation}
\langle \rho(\varepsilon) \rangle =
\frac{4|\varepsilon|}{\pi A} e^{-\frac{4\sqrt{2} \hbar
|\varepsilon|^{3/2}}{3mA}}, \label{resrho}
\end{equation}
which gives Eq.(\ref{K1}).

\section{First-order correction to the weight of the
optimal fluctuation of the electron-hole pair}
\label{appendixB}

The calculation of the first-order correction can be
performed in two different ways. On the one hand we can
use the exact relation between the action of the optimal
fluctuation and the disorder potential averaged over the
electron-hole wave function:
\[
\int\!\!d^2x \Psi^2(x_1,x_2) \left(U(x_1) - \beta
U(x_2)\right) = - \frac{1}{\lambda}\int\!\!dx U^2(x) = -
\frac{2}{\lambda} S.
\]
This equation allows us to write the action in the form
\[
S = \frac{\lambda}{2} \int\!\!d^2x \Psi
\left(-\frac{\hbar^2}{2m_h} \frac{\partial^2}{\partial x_1^2}
-\frac{\hbar^2}{2m_e} \frac{\partial^2}{\partial x_2^2} + V(x_1 - x_2) -
E \right) \Psi.
\]

The first-order correction to the action is then given by
\begin{equation}
S_1 = \frac{\lambda}{2} \int\!\!d^2x \Psi_0^2 V(x_1 - x_2)
+ S_1^{\,\prime}, \label{oneway}
\end{equation}
where
\[
S_1^{\,\prime} = \lambda \int\!\!d^2x \Psi_1
\left(-\frac{\hbar^2}{2m_h} \frac{\partial^2}{\partial x_1^2}
-\frac{\hbar^2}{2m_e} \frac{\partial^2}{\partial x_2^2} - E \right)\Psi_0
\]
Using Eq.(\ref{Psi0}) for the unperturbed electron-hole
wave function $\Psi_0$ and Eq.(\ref{nonlin}) for the
single-particle wave function $\phi$, we obtain
\[
S_1^{\,\prime} =  2 \lambda \int\!\!d^2x \Psi_0 \Psi_1
\left( |\varepsilon_h| \phi^2(\kappa_h (x_1-x_h)) +
|\varepsilon_e| \phi^2(\kappa_e (x_2 - x_e)) \right).
\]

On the other hand, the direct calculation of the
first-order correction to the action, in which one uses
Eq.(\ref{defaction}), gives
\begin{equation}
S_1 = - \frac{4 \varepsilon_h \varepsilon_e}{\beta}
\int\!\!dx \phi^2(\kappa_h (x-x_h) \phi^2(\kappa_e (x -
x_e)) + \int\!\!dx U_0(x) U_1(x), \label{another}
\end{equation}
where $U_1$ is the first-order correction to the disorder
potential:
\[
U_1(x) = - 2 \lambda \int\!\!dx' \left[\Psi_0(x,x')
\Psi_1(x,x') - \beta \Psi_0(x',x) \Psi_1(x',x)\right]
\]
The second term in Eq.(\ref{another}) can be rewriten in
the form
\begin{eqnarray}
\int\!\!dx U_0(x) U_1(x) &\!\!\!=\!\!\!& 4 \lambda
\int\!\!d^2x \Psi_0(x_1,x_2) \Psi_1(x_1,x_2)\nonumber\\
&\!\!\!\times\!\!\!& \bigg[ |\varepsilon_h|
\phi^2(\kappa_h (x_1-x_h)) + |\varepsilon_e|
\phi^2(\kappa_e (x_2 - x_e))\label{long}
\\&\!\!\!-\!\!\!& \beta^{-1}|\varepsilon_e|
\phi^2(\kappa_e (x_1 - x_e))- \beta |\varepsilon_h|
\phi^2(\kappa_h (x_2 - x_h))\bigg] \approx 2
S_1^{\,\prime},\nonumber
\end{eqnarray}
where in the last step, we used that the overlap between
the unperturbed electron and hole wave functions,
$\psi_h(x-x_h)$ and $\psi_e(x - x_e)$ is small, as a
result the third and the fourth terms in the square
brackets of Eq.(\ref{long}) give a much smaller
contribution than the first and the second terms. Thus, we
obtain
\begin{equation}
S_1 = - \frac{4 \varepsilon_h \varepsilon_e}{\beta}
\int\!\!dx \phi^2(\kappa_h (x-x_h)) \phi^2(\kappa_e (x -
x_e)) + 2 S_1^{\,\prime}. \label{anotherway}
\end{equation}
Combining Eqs.(\ref{oneway}) and (\ref{anotherway}), we
obtain
\begin{equation}
S_1 =  \lambda \int\!\!d^2x \Psi_0^2 V(x_1 - x_2) +
\frac{4 \varepsilon_h \varepsilon_e}{\beta} \int\!\!dx
\phi^2(\kappa_h (x - x_h)) \phi^2(\kappa_e (x - x_e)),
\end{equation}
which is equivalent to Eq.(\ref{S1}).

\end{document}